\documentclass[]{article}

\begin{document}

\title{The Origin of Semiconductor Physics in Italy: 1945 - 1965 \footnote{This
work has been carried out within the research project on the birth
of the physics of matter in Italy:
http://fisicavolta.unipv.it/percorsi/hip.asp}}

\author{Marco Luca Rossi}

\maketitle

\begin{center}
\emph{ Via Palmanova, 67 - 20132 Milano - Italy}
\end{center}

\begin{abstract}
The activities carried out by Italian physicists in the field of
semiconductor physics during the period spanning from 1945 to the
foundation of the National Group of Structure of the Matter (GNSM,
1965) are reviewed within their historical context. Until the
fifties, the Italian research was only marginally involved, if at
all, in the main streams of advancement in solid state physics.
Starting from the early fifties, an interest for  technical
applications of the newly introduced semiconductor devices began to
grow in the electronic engineering community. In the following
years, the birth of a few experimental and theoretical groups lead
by highly motivated scientists (some of them with international
experience)  allowed to deal with the main topics related to
condensed matter and semiconductor phenomena. The work developed by
these "pioneers", discussed in this paper,  represented an
invaluable contribution for the new generations of physicists in
this research field.
\end{abstract}

\section{Introduction}          
\label{sec:int}

The development of physics in Italy in the twentieth century has
drawn the attention of historians only rather recently. After the
pioneering work by Gerald Holton on Fermi's group \cite{holton},
several studies have dealt with institutional backgrounds \cite
{amaldi} \cite{russo} \cite{galda} \cite{cimento} \cite{cnr},
diffusion of relativity and quanta \cite{good} \cite{maiocchi}
\cite{mara} \cite{de} \cite{seba}, cosmic rays and nuclear physics
\cite{ivana} \cite{battimelli} \cite{orlando} \cite{reti}. The
attention has been mainly focused on the researches by and around
Fermi's group, while topics concerning what is now called the
physics of matter (atoms, molecules, liquids and solids) have been
investigated mainly (if not only) by Giuliani and his
coworkers\cite{ams}. The same is true for the post war years.
  \par
  The picture that emerges from these studies is now rather well
  established. Before the second world war, the Italian physical
  community was small  and scattered over the country: at the
  end of the thirties the number of academic physicists was around
  140 and the research units counted only few people. The
  smallness of the research groups
  made it difficult to follow the main streams of development of the
  discipline and/or master, at least, some of its up to date topics.
  The  experimental tradition inherited from the nineteenth century,
  in the absence of theoretical activity favored a
   negative reaction to relativity and quanta; the reaction
   with respect to quanta, in
   turn, hampered the entrance into  new
   fields of experimental research such as atomic and molecular
   spectroscopy. The physics courses maintained well into the thirties
   the structure and the contents inherited from the nineteenth
   century with only two novelties: the introduction, around 1920,
   of a course entitled
   Fisica Superiore (in which, but not everywhere, some
   relativity and   quanta topics were taught); and the introduction
   of a course of Theoretical Physics (Fisica Teorica) in the
   academic year 1936~-~1937\footnote{But, in that year, good courses
    of Quantum Mechanics were held only in four or five Universities \cite{cimento2}.}.
    Since then, we must wait till 1961 in order to see a new
    adaptation of the physics course to the advancements of the
    discipline with the introduction, in particular, of a course entitled
    `Struttura della Materia' (Structure of Matter).
    \par
    Since the thirties,  physical research in Italy is known
    because of the activities of Enrico Fermi and Bruno Rossi in the
    fields of nuclear and cosmic rays physics, respectively. But, as
    concerns the condensed matter, the scene was dim. During the first
    four decades of the century, experimental research  has been done
     in fields that later
    will be recognized as pertaining to solid state physics. The topics studied have been:
    a) elastic properties;
b) thermal properties; c) electrical properties; d) magnetic
properties e) galvanomagnetic and thermomagnetic effects; f) Volta
and photoelectric effects; g) optical properties of ions in solid
solutions or crystals. However, the application of
     Quantum
    Mechanics to the physics of crystalline solids
     (which started in late twenties)
     has been
    completely neglected in Italy \footnote{With some isolated exception: in late thirties,
    Giovanni Gentile jr. dealt with some problems concerning ferromagnetic properties.}. Therefore, Italian physicists
    working in these fields approached the end of the thirties with
    a cultural background typical of a pre~-~quantum era.
    \par
    The scenario Italy had to face after the second world war has
    been  described by Giuliani \cite{cimento3}:
`After the second world war, the international context of scientific
research appears profoundly changed: the defeat of nazism and
fascism and the war damages have favored the passage of the economic
and scientific leadership on the other side of the Atlantic; the
huge effort by the United States in the production of the fission
bomb has shown how efficient can be a research based on a planned
mixing of basic, applied research and technology; the shock provoked
by the nuclear bombing of Hiroshima and Nagasaki has again
emphasized the necessity of a reflection about the aims of Science
and its applications. The start of the cold war and the consequent
search of new weapons has dramatically increased  the interest of
governments in the military and peaceful applications of the new
technologies and has spurred the development of research programs of
unprecedented economic and organizational commitments, often
affordable only through international cooperation. The problems that
Italy had to face in order to achieve the scientific and
technological background necessary for the country's development
were complex and difficult: the war damages and the overall weakness
of scientific institutions made the task even more arduous.'
\par Also the task of those researchers who, for personal choice
and/or local influences began to work on the physics of condensed
matter, was arduous. They have inherited a cultural background in
which the physics of condensed matter was still that of the
twenties; the university courses they have followed scarcely
contained Quantum Mechanics; even less, its application  to
condensed matter topics. They had to face a knowledge gap of about
twenty five years.
\par
Condensed matter research developed in Italy as a polycentric
process in which small local groups began to work with more contacts
with foreign groups than among themselves \cite{ggib}. In this
context, particularly interesting has been the birth of
semiconductor physics. Ma\-raz\-zi\-ni and the author of this paper
have recently published a book on this subject \cite{rossi}. This
paper aims at making available to a wider audience the most
significant results of that research.

\section{The background}          
\label{sec:con}

In his opening lecture at the ``Solid State Physics School'' held in
Varenna in 1957, Fausto Fumi defined solid state physics as ``...
the study of the physical properties of solids and of the particular
properties exhibited by atoms and molecules as a consequence of
their reciprocal interactions and their regular distribution within
the lattice''\cite{Fum-var}. Actually, the two main subjects dealt
with were the Quantum Theory of Solids and the Lattice Defects,
whereas topics regarding semiconductors  were
 considered as a ``product'' deriving from these more general
studies.

The delay of  Italian research in solid state physics is well
documented by the program of the  Meeting of the Italian Physical
Society (SIF) held in Como in 1947: it was the first SIF Meeting in
the post-war period. Antonio Carrelli's talk, ``Modern aspects of
physics research'', was about the same subject dealt with in Como
twenty years before: the application of Fermi statistics to the
 theory of metals by Sommerfeld.  Twenty five years later, Luigi Giulotto
 stressed   that the only significant
studies performed by Italian physicists in solid state physics until
the 1950s have been the studies (in the thirties) on the Raman
effect by Carrelli in Naples and by Rasetti, Amaldi and Segr\`{e} in
Rome (modestly ignoring his own contributions in Pavia)
\cite{giu1982}.

In order to shed some light on the main bottlenecks that have
hampered the development of   solid state research, it is worth
discussing in some detail: i) the prevailing role of Nuclear
Physics; ii) the connection between academic institutions and
industry; iii) the distribution of  funds; iv) the small number of
research groups and the lack of coordination among them.
\\
\\
i) From the thirties  to the end of the Second World War,  a
significant progress has been made in the understanding of
semiconductors properties. The post-war years have been
characterized by an increasing  interest in the field, especially
after the invention of the transistor in 1947. However, in Italy,
the situation was rather different.

Antonio Rostagni opened the  SIF Meeting in 1948 with a lecture
entitled ``Recent discoveries and new instruments in Physics
research'', that was almost entirely dedicated to Nuclear Physics
and  instruments such as synchrotrons and photomultipliers
\cite{rost}. In particular, no mention of the invention of the
transistor was made.

At the  SIF Meeting in 1961, Giampiero Puppi (engaged in particle
Physics research) acknowledged  that the funds devoted to elementary
particles Physics were far greater than the ones allocated to other
Physics fields \cite{Pup-conf}. According to Puppi, this situation
has been caused by the exceptional development of  Fermi and  Rossi
schools during the 1930s. However, instead of asking for a more
equilibrated allocation of funds, Puppi held that the way towards a
better equilibrium would have been to stimulate the choices of young
physicists towards less developed fields.

Six years later,  Luigi Giulotto underlined the lasting of this
unequal distribution of funds in a well  documented paper published
in the bulletin of the National Group of Structure of Matter (GNSM)
\cite{giu-gnsm}.
\\
\\
ii) Industries and academic Insitutions have been (almost)
completely separated till the sixties. This is  particularly true
for semiconductor physics. As Gianfranco Chiarotti put it at the SIF
Meeting in 1962 ``\dots the Italian research in the field of
semiconductors is practically missing [while] the main interest
\dots  seems to deal with  the study of lattice defects\dots''.
According to  Chiarotti, the main causes were ``\dots the lack of a
robust scientific tradition [in this area], the absence of a good
education system in the field of solid state physics and the almost
complete lack of interest of Italian industry in  scientific
research''\cite{chi-nota}.

In  the late 1950s the first Italian semiconductors firms have been
founded: SGS (Societ\`{a} Generale Semiconduttori) in Milan and ATES
(Aziende Tecniche Elettroniche del Sud) in Catania. The activities
of these two companies were focused on the production of electronic
devices. The cooperation between industries and Universities was
limited to the
 evaluation of some physical parameters and to the
collaboration with  few students working on their thesis.

In 1963 a very small percentage (9\%) of Physics graduates were
employed in industry, whereas most of them worked in academic
research (78\%) or in other sectors (13\%) \cite{sac}. This trend
has been confirmed by Guido Tagliaferri in 1964: see Table
\ref{tabtaglia} \cite{tagl}.
\begin{table}
\caption{Occupational distribution of Physics graduates
(Ref.\cite{tagl}).}
\label{tabtaglia}                    
\begin{tabular}{lrr}
\hline\noalign{\smallskip} Category\\ of employment & Number &\% \\
\noalign{\smallskip}\hline\noalign{\smallskip}
Research institutions & 494 & 64,5 \\
Industry  & 70 & 9,2 \\
Commercial sector& 16 & 2,1 \\
Teaching& 103 & 13,5 \\
Others& 8 & 1,0 \\
Waiting military service & 25 & 3,3 \\
Unemployed  & 49 & 6,4 \\
\textbf{Total}  & \textbf{765} & \textbf{100,0} \\
\noalign{\smallskip}\hline
\end{tabular}
\end{table}
See also Giulotto's comments on Tagliaferri data  \cite{giu-gnsm}
\footnote{As recalled by Giuliani, ``Starting from the sixties,
Giulotto began a long lasting activity aiming at an equilibrated
development of physical researches: the Physics of the Matter, whose
technological outcome was of strategic relevance, was badly
underdeveloped. However, Giulotto commitment, though based on a
correct analysis and sustained by an ambitious goal - to obtain what
will be later called INFM (National Institute for the Physics of
Matter) - has been hampered by tactical errors and resentful
polemics. The path has been difficult and perilous: as a matter of
fact, the INFM was founded only in 1994.'' \cite{giubiogiu}.}.
\\
\\
iii) At the  SIF Meeting in 1962, the Minister of Public Education
(Pubblica Istruzione), Giovanni Medici,  stated that the budget for
scientific research was about 50 billions  lire per year, equivalent
to 0.2\% of the gross domestic product, whereas in other countries
the total research investment was ten times greater \cite{medi}.
Medici also acknowledged that the funds were often distributed
without considering the actual needs of the different lines of
research.

A detailed analysis of the financial support of  solid state physics
in  late 1960s was published by Chiarotti in 1982 \cite{chia-1982}.
Chiarotti recalled that in the years 1965, 1966 and 1967 the yearly
investment of CNR (Consiglio Nazionale delle Ricerche) was greater
than 20 billions Italian lire, but less than 4\% of this amount was
allocated to solid state physics research.
\\
\\
iv) The problems arising from the small number of  Solid State
research groups have been  highlighted by Giulotto at the solid
state physics School  held in Varenna in 1954. Giulotto stressed
that the main purpose of the School was to give the small Solid
State community the possibility of discussing the most relevant
topics of their discipline and their own results with a few
international specialists\cite{giu-var}. In order to better grasp
the real meaning of the word ``small'' used by Giulotto, it is worth
recalling that at the SIF Meeting held in the same year only one
talk (presented by Fausto Fumi) was on solid state physics.
\par
During the SIF Meeting  held in Bari in 1963, a group of
 researchers founded a voluntary association called
``Gruppi Italiani di Struttura della Materia'' (GISM) whose main
purpose was to promote the research  in the field of atomic,
molecular and solid state physics. This organization was formally
recognized by CNR in 1965  as the ``Gruppo Nazionale di Struttura
della Materia del CNR'' (GNSM): eleven years later, the GNSM counted
24 research groups. In the same year  the first issue of the already
quoted GNSM bulletin was published. Its goal was the sharing of
``\dots information and comments about the progress in an area of
Physics that has been characterized by a dramatic development, that
is Condensed Matter Physics: atoms, molecules, solid  and liquid
state of matter''.  In a recent recollection of those years,
Chiarotti stressed how ``\dots in the 1960s the GNSM was a group of
people driven by the same motivations (\dots) The intellectual
commitment was extremely high and there was also the strong feeling
that the researches [in that field] would have had high relevance
for the Italian research and for the country''\cite{chia-pv}.

\section{Engineers before Physicists}
\label{sec:eng}

In Italy,  semiconductors topics  have been firstly addressed to by
the electronic engineering community. The first  Italian reference
to transistors appeared in a short note published in the magazine
``Poste e Telecomunicazioni''  in 1949\cite{poste-tel}. This note
described the  structure of the point contact transistor, the
polarization details, the amplification properties and the
equivalent circuit.

In 1950, the same magazine published other papers focused on the new
semiconductor device, written by Piero Schiaffino \cite{schiaf} and
Lamberto Albanese \cite{alba}. Other papers appeared  in  magazines
such as ``Alta Frequenza'' and ``L'Elet\-tro\-tec\-ni\-ca'' and
again in ``Poste e Telecomunicazioni''\footnote{A complete review of
these publications can be found in Ref.\cite{rossi}.}. These papers
dealt
 with the technological and circuital aspects of
semiconductor devices. Among other contributions, it is worth
recalling the experimental studies  by Manfrino
\cite{man1}\cite{man2}\cite{man3}, Sette and Della Pergola
\cite{set1}\cite{set2} working at the ``Istituto Superiore di Poste
e Tecomunicazioni'', with the support of the ``Fondazione Bordoni''.
These papers, as well as the one  by Della Pergola and Sette
published in ``Il Nuovo Cimento'' in 1956 \cite{perg-set-1}, were
the first focused on the physical properties of  semiconductor
devices.

\section{The first developments}
\label{sec:fdev}

The  late 1950s showed a growing interest in semiconductor physics:

a) in 1956, Franco Bassani began to publish several works on band
structure calculations;  b) in 1957, the School of solid state
physics in Varenna included a special section on semiconductors; c)
in 1957, the already mentioned  SGS (Societ\`{a} Generale
Semiconduttori)  began to produce semiconductor devices under
Fairchild patents.

A quantitative description of the growing interest in semiconductor
physics is given in Table \ref{tab:2}: the numbers refer to papers
published  by Italian physicists in the following magazines:
\begin{itemize}\small
\item Alta Frequenza \item Il Nuovo Cimento \item Journal of
Applied Physics \item Journal Phys. Chem. Solids
\end{itemize}
\begin{itemize}\small
\item Philosophical Magazine \item Physical Review \item Physics
Letters \item Physics Review Letters
\end{itemize}
\begin{table}
\caption{Number of Italian papers on semiconductors in the period
1956 - 1966 published in the magazines listed in the text
(Ref.\cite{rossi}).}
\label{tab:2}                    
\begin{tabular}{lcc}
\hline\noalign{\smallskip}
Year & Theoretical & Experimental\\
\noalign{\smallskip}\hline\noalign{\smallskip} \hline
1956&1&3\\
1957&2&2\\
1958&0&1\\
1959&4&1\\
1960&0&0\\
1961&4&4\\
1962&4&3\\
1963&4&5\\
1964&0&10\\
1965&1&6\\
1966&3&11\\
\textbf{Total}&\textbf{23}&\textbf{46}\\
\noalign{\smallskip}\hline
\end{tabular}
\end{table}
The data reported in Table \ref{tab:2} can be compared with those of
Table \ref{tab:3} concerning  the communications presented by
Italian scientists at the SIF Meetings in the same period.
\begin{table}
\caption{Italian communications at the SIF Meetings on
semiconductors in the period 1956 - 1966 and their percentage with
respect to the total number of communications (Ref.\cite{rossi}).}
\label{tab:3}                    
\begin{tabular}{lcc}
\hline\noalign{\smallskip}
Year & Number & \% \\
\noalign{\smallskip}\hline\noalign{\smallskip} \hline
1956&1&2,8\\
1957&-&-\\
1958&3&3,0\\
1959&1&1,6\\
1960&0&1,3\\
1961&2&2,0\\
1962&1&5,3\\
1963&3&5,8\\
1964&5&5,1\\
1965&7&5,8\\
1966&5&4,3\\
\textbf{Total}&\textbf{28}&\textbf{4,0}\\
\noalign{\smallskip}\hline
\end{tabular}
\end{table}
\par
In 1965 Sette published a \emph{Quaderno} on  the free electron
theory in metals, the energy bands theory and  semiconductor
properties \cite{quaderni}. In the same year, Giampaolo Bolognesi
published the volume \emph{Tecnologia dei semiconduttori}: after a
short introduction on semiconductor theory, the book dealt with
semiconductor technological processes \cite{bolognesi}. The rich
bibliography of this volume did not quote any Italian contribution.

\section{Experimental researches}         
\label{sec:exp}

\subsection{In Rome.} The first Italian experimental researches on  semiconductor
properties have been carried out at the ``Istituto Ugo Bordoni''.
This institution was founded within the ``Istituto Superiore delle
Poste e delle Telecomunicazioni'' in 1952.

Ten laboratories were supported by the ``Istituto Ugo Bordoni'' and
their activities dealt with several topics, such as: microwaves,
semiconductors, wave propagation, electroacoustics, electronic
microscopy, transistor based circuits and radio communications. The
semiconductor laboratory was directed by Daniele Sette
\cite{bib_book}:  it represented the main Italian
 activity in that research area in the 1950s. The
scientists working with Sette  were Renato Manfrino, Giancarlo Della
Pergola and Venanzio Andresciani.

In 1954, Sette spent a short period in the USA and France in order
to gather information about the most recent  lines of research in
the field of semiconductors. In the USA, he visited several
institutes and attended the Conference of the American Acoustic
Society and the Conference of the Institute Radio Engineers, both on
semiconductor topics. He then visited the French laboratories of the
Centre National d'Etudes des Telecommunications (CNET).

This journey allowed Sette to get in touch with experimentalists and
theorists working in fundamental semiconductor research. He  also
 established a communication network providing
information about research in progress and available post-doc
positions. Sette identified a list of  the  main techniques used to
study semiconductor properties. This list became the agenda on which
he decided to concentrate the activity of his laboratory
\footnote{Daniele Sette, private communication.}. Various
experimental setups were developed and used in galvano~-~magnetic
measurements (Hall effect and magneto~-~resistance) and in
photoconductivity experiments.  Among these studies: a work on
chemical attacks on Germanium crystals
 \cite{perg-set-1};  a study on
 the ratio  between
the longitudinal and transversal effective masses in Germanium
\cite{{perg-set-mag}};
a paper  on the experimental verification of the theoretical
hypothesis  concerning the energy band structure in Germanium and
Silicon \cite{perg-set-ger} \footnote{In this work, the available
information (obtained by cyclotrons resonances measurements) on the
band structure of Germanium was completed by measuring the Hall
coefficient of \emph{n} and \emph{p} $Ge$ as a function of the
applied magnetic field. The results obtained were in agreement
 with the ones published in the same period by
Goldberg and Davies \cite{gold-davies}.}; and, finally, the study
on the mean lifetime of charge carriers \cite{bed-set}. In this
work, the mean lifetime was ultimately obtained by using the theory
developed by Shochley \cite{shockley} and Rittner \cite{rittner}.

 A detailed discussion of the research activities carried out by
 Sette and coworkers
 can be found in  \cite{rossi}.

\subsection{In Cagliari and Bologna} \label{sec:exp1}  Semiconductor
studies in Cagliari have been promoted by Giuseppe Frangia. In 1959,
he offered to Pierino Manca (graduated in Chemistry) a research
position for the study of semiconductor properties. At that time,
Frangia's staff was composed by Carlo Muntoni, Francesco Aramu and
other young students. The first activity of this group concerned the
preparation of CdS single crystals samples using the  the vapor
phase growth technique \footnote{Frangia's group introduced it in
Italy.}.

 Manca and his co-workers
studied the semiconducting properties of Iron Telluride
\cite{ara-tel} and the photocurrent decay processes in
polycrystalline CdS \cite{ara-phot}.

Manca worked on Sulfide compounds until 1965. In that year, together
with Aramu, he studied the crystallographic structure as well as the
electrical, magnetic and optical behavior of these compounds. The
know-how acquired during the experimental studies on Tellurides
allowed Manca and Massazza to prove that the properties of
AgFeTe$_{2}$ could be related to the existence of an equimolecular
mixing between Ag$_{2}$Te and Fe$_{2}$Te$_{3}$\cite{man-mass}
\cite{man-mass-1}.

Another line of  research concerned the optical properties of
semiconductors. These studies started in 1962, when Frangia assigned
a thesis on the luminescence of CdS to Francesco Raga. The first
work appeared in 1963 \cite{man-rag-nat}: it concerned the
absorption band edge in wurtzite type CdS before and after the
structure transition CdS$_{Z}$ (Cadmium Sulfide with zinc-blende
structure) $\rightarrow$ CdS$_{W}$ (Cadmium Sulfide with wurzite
structure) and its shift as a function of temperature.

The studies about the preparation of solid compounds
CdS$_{x}$Se$_{1-x}$ ($0 \leq x \leq 1$) yielded single crystals that
were used in optical studies. The luminescence spectra of these
materials were analyzed  and the results  compared with the ones
obtained by thermoluminescence techniques. The data obtained
confirmed the correlation between the change of the energy gap and
the shift of the upper limit of the valence band, while the lower
limit of the conduction band remained unchanged. Moreover, the
authors were able to shed light on the observed change in the
exciton luminescence spectral line. In the following period Raga
kept working on these luminescence phenomena in collaboration with
Nikitine and others \cite{kleim} \cite{mysy} \cite{mysy-grun}
\cite{mysy-grun-biv}.
\\
In 1964, Manca, Muntoni and Raga published a paper on the behavior
of exciton states under thermal gradient \cite{man-mun-rag}. This
subject was extensively investigated by Raga and Nikitine in the
following years \cite{kleim} \cite{mysy} \cite{mysy-grun}
\cite{mysy-grun-biv}. \vskip5mm
\par
In  late 1960s, other experimental groups investigated metal and
semiconductor properties using different techniques. In the field of
electron microscopy, we must recall the activity carried out
 in Bologna. In the 1950s, the Physics Institute was directed by Ugo
Valdr\'{e},  who introduced the electron microscopy technique in
Italy. In  early 1960s, Valdr\'{e} invited to Bologna Primo Gondi, a
young researcher with a good know-how on structural properties of
metals. The main topic investigated, by electron microscopy
technique, concerned   dislocations in Germanium crystals and their
correlation with plastic deformation, crystal hardening and thermal
treatments \cite{g-me}\cite{valdre}.

\subsection{In Pavia and\dots} As already mentioned, the international growth of solid state
physics was exceptional during the 1950s and the 1960s. The
flourishing of solid state physics, and more specifically of
semiconductor studies, was also stimulated by the possible transfer
of basic research into  commercial and military products, such as
semiconductor devices. The  production of these devices presented
various problems  mainly related to  minority carrier currents in
diodes, the  excessive noise, the surface recombination of minority
carriers and the different semiconducting characteristics of several
alloys. These problems stimulated a series of experimental and
theoretical researches on
 surface states in semiconductors.

In Italy, the experimental side of this field was deeply
investigated by a group of researchers at the University of Pavia
supervised by Chiarotti \cite{chiarotti}. Chiarotti's team was
composed by Adalberto Balzarotti, G. Del Signore, Andrea Frova,
  Giorgio Samoggia and Angiolino Stella: it began his
activity in 1960.

In that year Chiarotti and his co-workers presented a communication
at the SIF Meeting entitled ``Surface conductivity measurements in
Germanium'' \cite{chiar-ger}. The Hall coefficient, the conductivity
and the mobility of  surface carriers were measured in a
semiconductor whose surface  has been exposed to the so-called
Bardeen ~-~ Brattain cycle \cite{bratt-bard}. One year later, Frova
and Stella published a paper on  Tamm theory of surface states
\cite{frov-stel}. This paper dealt with the details of surface
states theory; it emphasized the existing contrast between
theoretical predictions and experimental results regarding the
change of surface potential produced by an electric field applied
perpendicularly to the surface of the sample. On the basis of these
considerations, Frova and Stella underlined the hypothesis of a
strong correlation between  surface level structure and chemical or
mechanical treatments.

The low density of surface states made it very difficult to detect
optical transitions from occupied valence band states to unoccupied
surface states (or from occupied surface states to unoccupied
conduction band states). In 1962, Harrick published a paper in which
he described the field effect modulated optical absorption technique
\cite{harr}: the use of a sinusoidally varying electric field made
it possible to detect, by looking at the modulated absorption, the
optical transitions to or from  low density electronic states. In
the same year, Chiarotti's group began to use  this technique for
studying semiconductor surface properties. The first paper  appeared
in 1962  and dealt with   the effects on Germaniun surface produced
by Oxigen \cite{chi-sig}. The dependence of fast surface states on
surface contamination was further examined by Chiarotti, Frova and
Balzarotti \cite{balz-fast} by exposing a sample of Germanium  to
different environmental conditions. Some technical improvements
apart, the experimental method used  by Chiarotti and his co-workers
was quite similar to the one discussed by Bardeen et al. in a  work
published in 1956 \cite{bardeen1956} \footnote{The technique was
based on the measurement of the conductivity as a function of the
surface potential modulated by an electric field applied
perpendicularly to the surface.}.

In 1962 Chiarotti obtained a permanent position at the University of
Messina, where he moved in the same year together with  Balzarotti,
Frova, Umberto Grassano, Andrea Levialdi and Gianfranco Nardelli.
However, Chiarotti did not  stop his collaboration with the rest of
the original group in Pavia: a paper focused on the optical
detection of surface states using the modulated field effect
technique  was published in 1966 \cite{snc-od}. The main result  was
the detection of energy levels correlated to a particular transition
from surface states to the conduction band.

 In 1963,  Frova accepted a Research Associate position
at the Semiconductor Research Laboratory in Urbana (Illinois),
directed at that time by John Bardeen. This position has been
offered by Paul Handler, working at the same lab as vice-director.
After joining Handler's group, Frova began to apply the
electromodulation spectroscopic techniques used in Pavia to  p-n
junctions in Germanium,  Silicon and  Gallium Arsenide. As a matter
of fact, this experimental setup had been previously used by
Chiarotti in Pavia in the study of the Franz-Keldysh effect
\cite{Keldysh} in semiconductor surfaces. In 1965, Frova and Handler
carried out an experimental study of the Franz-Keldysh effect in a
p-n Germanium junction \cite{fh-es}.

Using  the same technique, Frova and Penchina measured  the energy
gap in Germanium obtaining values in good agreement with the ones
obtained by  the more expensive and complex methods based on
photoconductivity and magneto-absorption processes \cite{fp-eg}.
 Frova, Handler and
co-workers, studied also the variation of the optical absorption
coefficient in Silicon and Germanium \cite{fhga-ea}. This paper
dealt with a
 detailed study of the direct and indirect
(phonon assisted) intraband transitions in both elements.

In the same period other investigation methods were developed by
various researchers based on electroreflectance and wavelength
modulation. In the following years, all these experimental
procedures, referred to as \emph{modulation spectroscopy
techniques}, allowed to obtain a great deal of information on
vibrational and electronic properties in solids.

Another member of Chiarotti's former group was Angiolino Stella. He
spent a few  years  at the Iowa University (from 1961 to 1964) to
carry out studies on semiconductor compounds. Stella published
several papers in collaboration with D.W. Lynch on this subject
\footnote{Lynch has been in Pavia in 1959, working with Chiarotti on
 color centers in KCl.}. We recall here the first one concerning
the semiconductor compounds Mg$_{2}$Si and Mg$_{2}$Ge \cite{sl-pm}:
significant physical parameters, such as the energy gap and the mean
life  time of charge carriers, were determined. These kind of
studies, focused on II-IV compounds, represented the core of the
activity carried out by
 Lynch and Stella until the late 1960s.

Finally, it should be mentioned  the paper on the pressure
coefficient of the band gap in  Mg$_{2}$Si, Mg$_{2}$Ge and
Mg$_{2}$Sn\cite{sbhl-pc}:  the experimental results yielded
information about the structure of band edges, despite  the low
quality of the samples and the low pressures used.

\subsection{Researches in non academic Institutions} We shall briefly discuss the activities developed at the Olivetti
laboratories located in  Milan area. The first communications by the
Olivetti group appeared at the SIF Meeting in 1963.
Franco Forlani and Nicola Minnaja, focused their attention on the
investigation of semiconductor structures related to planar
microelectronic devices.

In 1964 Forlani and Minnaja published a paper, ``Conduction
Phenomena in Si$-$SiO$_{2}$$-$Al Structures'', on the properties of
the electrical conduction in semiconductor~-~dielectric~-~metal
structures \cite{fm-cp}. For a closer look at the research activity
of both authors one can see the Proceedings of the SIF Meetings of
those years \footnote{An stimulating recollection by Forlani about
the events of those years can be found in Ref.\cite{rossi}.}.

Other Institutions contributed to semiconductor research in those
years. Here, we simply enumerate their fields of research:

\begin{itemize}
\item[-] CISE: studies on
Silicon detectors.

\item[-] EURATOM (Ispra): studies on the
so-called forbidden reflection in Silicon and Germanium.

\item[-] Istituto Galileo Ferraris (Torino): studies on
piezoresistance effects in n-type  semiconductors.

\item[-] Istituto Superiore di Sanit\`{a}: studies
on Germanium with special reference to  the complex dielectric
constant in  intrinsic Germanium in the microwave region.
\end{itemize}

>From 1961 to 1966, several groups working in different institutions
developed experimental researches on radiation effects in
semiconductors. The first work on this subject was presented at the
SIF Meeting in 1959 by G. Airoldi, Z. Fuhrman, and E. Germagnoli and
published in ``Il Nuovo Cimento'' in the same year \cite{afg-cr}
(conduction properties and Hall effect in electron irradiated
Germanium).  As concerns neutron radiation effects in
semiconductors,
 Mario Bertolotti and Sette published the first
Italian paper on these subjects in 1961 \cite{bs-rp}. In their
study, the mean life time of minority carriers in  irradiated
Germanium was analyzed in order to verify some hypotheses proposed
by Crawford \cite{cr-gos1} and Gossick \cite{cr2}.
 The activity of Sette's group on neutron
irradiated  semiconductors was very lively in the first half of the
Sixties as shown in \cite{bpsgv-od} \cite{bpsv-em}.

Radiation effects with alpha particles (Colella and Merlini at
EURATOM) and ionized O$_{2}$$^{+}$ molecules (Bacchilega, Gondi and
Missiroli in Bologna) were extensively analyzed during the period
1964~-~1965.

Another research topic was about positrons produced by gamma rays in
semiconductors. The first experimental results have been described
in two papers published in 1963 and 1964 by P. Colombino, B.
Fiscella and L. Trossi \cite{cft-ps} \cite{cft-ad}. The first paper
presented the detailed description of the linear slits and the point
slits methods \footnote{This technique was originally developed by
S. De Benedetti and L. G. Lang in order to study amorphous
polycrystalline substances\cite{dl}.}. The second paper concerned
instead the angular distribution of annihilation quanta emerging
from Si, Ge and Al crystals. The positron decay process in Silicon
was also studied
 at CISE laboratories. 
These studies aimed at determining what kind of electrons (valence
or conduction) was involved in the annihilation process. Later,
 the positron mean lifetime in
different semiconductors, such as Gallium, Silicon, Boron and
Silicon Carbide has been studied.
\\
\\
>From a more technical viewpoint, the two areas of main interest were
the crystal growth and the improvement of measurement techniques.
The production of semiconductor samples constituted a vital issue.
This is evident if one considers that the Germanium samples used in
the laboratories of the ``Istituto Ugo Bordoni'' came from the USA
until the late 1950s. Furthermore, in the early 1960s a great number
of Italian researches have been carried out thanks to samples
donated by S.G.S., the Italian semiconductor manufacturer founded in
1957. Several papers on semiconductors sample preparation have been
published in the magazine ``Alta Frequenza'' by Manfrino, Sette,
Della Pergola and Venanzio Andresciani in the late Fifties
\cite{perg-set-att}\cite{m-ps}\cite{as-sa}. In Ref.\cite{as-sa},
Sette and Andresciani  discussed the main characteristics of a
furnace (Czochralski type) used to prepare Germanium crystals of
very high quality.

In the field of electronic microscopy, the review  by Gondi in 1965
\cite{g-me} should be mentioned. It shows the enduring interest  in
the application of this technique to the study of dislocations and
lattice defects in semiconductors.

An apparatus for differential spectroscopy has been developed  by
Guido Bonfiglioli and Pietro Brovetto (both working at the
``Istituto Galileo Ferraris'' in Turin)\cite{bb-ps}
\cite{bbblpw-sm}. This instrument, of wide application, has been
used also in semiconductor research.
\\
\\

\newpage

\section{Theoretical researches}          
\label{sec:theo}

With the exception of two reviews published by Antonio Carrelli in
1946 and in 1949 on the Hall coefficient in Bi$-$Sb and Bi$-$Te
\cite{carr1} \cite{carr2}, the Italian contribution to the
theoretical research on semiconductors was completely absent until
the mid Fifties.

In  early Fifties the interest for the theoretical aspects of solid
state physics began to grow in the Italian community, mainly thanks
to the efforts of Piero Caldirola  \cite{caldbio} (then director of
the Physics Institute in Milan) who put in contact some students
coming from Pavia University (Franco Bassani, Roberto Fieschi and
Mario Tosi) with Fausto Fumi\cite{fumibio}, a young theorist who had
spent some time in the U.S.A. working with Frederick Seitz at the
University of Illinois.

The  role played by Fumi in the birth of solid state physics in
Italy
 (not only in theoretical research), is
underlined by Chiarotti in a recent recollection:``In the summer of
1951 I was doing my thesis under L. Giulotto (...) I attended an
informal seminar by Fausto Fumi (...) Fumi was talking of defects
showing, to our surprise, that many properties of solids depend more
on defects than of the regularity of the lattice. The subject of
seminar was the color centers, a thing we had never heard about:
Fumi discussed the models proposed for the various centers, models
that could be verified by simple spectroscopic experiments. Giulotto
grasped immediately the interest of the field but even more the
possibility of exploiting the remarkable optical equipment possessed
by the Institute. (\dots) So he asked me to work part-time on the
problems Fumi had suggested...''\cite{GiuChi}.

The experimental and theoretical study of defects in solids
represented a fundamental step in the development of  Italian
 solid state physics.
As Chiarotti put it at the SIF Meeting in 1963 ``...the Italian
interest seems to be focused on the properties related to the
presence of lattice defects (52\% of the total production in Italy
and 15,1\% in other countries)...''. A summary of the Italian works
on  lattice defects and color centers can be found in
Ref.\cite{rossi}. In the following the discussion will, of course,
be concentrated on the  studies carried out by Italian physicists in
research areas more strictly related to semiconductors.

In his communication at the SIF Meeting in 1963, Franco
Bassani\cite{bassbio} pointed out that ``\dots the number of Italian
physicists studying condensed matter is rather small and an analysis
of their contributions can not give us a complete idea of the
importance of this field (\dots) nevertheless in the last two years
the Italian contributions have been so important and
significative...''\cite{bass-rec}. Bassani listed the main three
subjects studied by Italian theorists: i) interatomic forces,
thermodynamic functions in ionic crystals and in rare gases; ii)
lattice defects; iii) electronic levels in metals and insulators. In
the following section we shall focus  on the latter subject.

\subsection{Electronic levels of solids}
\label{sec:theo1} A  review of the development of the band theory of
solids can be found in Ref.\cite{OCMaze}, where the most important
calculation techniques of electronic levels are discussed in their
historical background. In 1937, during his sabbatical semester at
the Princeton Institute for Advanced Study, Slater developed the
so-called Augmented Plane Wave method (APW) employing a
``muffin-tin'' potential. Two years later, Herring, a postdoctoral
fellow at MIT, elaborated what became known as the Orthogonalized
Plane Wave method (OPW), where the wave function is a linear
combination of low-lying atomic orbitals and plane waves. By the
mid-1950s, Solid State theorists had several methods at their
disposal that could be used to solve the Schrodinger wave equation
in a periodic potential. However, until the late Fifties the band
structure of solids was  considered  too difficult to be calculated.
Nevertheless, a few rough computational works had been performed in
the Forties in the USA using electromechanical computers.

At the International Conference on Semiconductors in Rochester in
1958, the only two communications on  energy band topics have been
presented by J.C.Phillips and G.F.Bassani but, as the Italian
scientist recalls, those sessions were attended only by few
participants\footnote{Franco Bassani, private communication.}.  At
that time, Franco Bassani was working at University of Illinois in
Urbana (1954-1957).

Bassani  recalled that during his stay in Urbana  ``...Fred Seitz
had given a course on group theory; the book adopted was a
photostatic copy of Wigner's book ``Gruppentheorie und
Wellenmechanik'' (\dots) we could learn the usefulness of symmetry
in defining general properties of quantum states. F. Seitz had
previously given a course on the electronic structure of solids, and
had invited Truman Woodruff to give a lecture on the Orthogonalized
Plane Wave Method which was just beginning to be used by F. Herman
and T.O. Woodruff''\cite{chiar-seitz}.

The studies carried out by Bassani in the late Fifties concerned
mostly pseudo~-~crystals, i.e. ideal crystals composed by atoms
which are known to crystallize in some other lattice. This area of
research was  suggested by Seitz as an useful tool for  studying the
influence of lattice symmetry and atomic potential on band
structure.

In 1957 Bassani published  an OPW calculation of the energy band of
Silicon \cite{bass-1} without using any ad hoc parameters.  Bassani
showed that the energy band structures are mainly determined by the
lattice symmetry properties. At the SIF Meeting in 1958,  Bassani
presented two papers on this subject. In the first communication,
Bassani and Celli analyzed the energy band structure of lithium
atoms in the diamond lattice, and after comparing their results with
those obtained by Herman \cite{herm} they observed that ``...the
order of the energy bands is mainly determined by the lattice
symmetry...'' In the second contribution, Bassani and Ignazio Fidone
described the energy band structure of 2-dimensional graphite by
using the Tight-Binding method. They
 concluded that ``\dots graphite is a borderline case between
metal and semiconductor as the occupied  and the empty states are
very close\dots''.

The theoretical works developed by Bassani and Celli on
pseudo~-~crystals deserve to be analyzed in more detail as they
introduced a powerful and simple perturbation approach, based on the
OPW method. They  considered the crystal potential as a small
perturbation on an empty lattice whose states were classified on the
basis of the crystal symmetry. From a technical point of view, the
fact that this method could give a reliable picture can be justified
by the ``cancellation effect'' due to the orthogonality condition
between the electronic wave functions and the core states. This
effect can be physically described as a repulsive term that
partially cancels the crystal potential. Such a formulation allows
to consider a total effective potential referred to as
``pseudo~-~potential''.

The same theoretical approach was used for studying  the energy band
structure of Lithium  \cite{basscell2}.  Bassani and Celli pointed
out that ``...the  approach leaves uncertain some of the detailed
features of the energy bands such as the curvatures, but we have
preferred not to use any interpolation, since we are mainly
interested in the sequence of the energy bands and in their
qualitative features''.

The energy band structures in pseudo~-~crystals were further
investigated by Bassani \cite{basscell3}: he studied the most
important factors determining the energy band structure of a
pseudo~-~crystal made of Sodium atoms in a diamond lattice by using
the OPW method. The lattice parameter of the pseudo~-~crystal was
chosen in such a way as to give the same  electron density as in
metallic Sodium. Bassani stressed that ``...it seems possible to
conclude that the sequence of the energy bands is mainly determined
by the lattice symmetry and their separation by the electron
density''.

In 1961, Bassani and Celli wrote a lengthy paper on the
pseudo~-~potential technique \cite{basscell4}. They stressed the
role of the perturbation approach and applied the technique to some
crystals with the diamond structure. For Germanium and Gallium
Arsenide the results obtained were in good agreement with the
experimental data of Kleinman and Phillips \cite{klph}. The authors
pointed out that, on the basis of the applied method,
  an increase of the
lattice constant entails a contraction of the energy scale of the
empty lattice eigenvalues making the separation between the final
energy levels smaller. This prediction was in agreement with the
experimental observation that the energy gap and the valence band
width decreases in going from diamond to Silicon and  to Germanium.
On this basis, Bassani and Celli explained  qualitatively the effect
of pressure on the energy gap: an increase in pressure caused a
decrease in the lattice constant and then an increase in the energy
gap, as experimentally observed. Finally, Bassani and Celli
underlined that  ``\dots this approach (\dots) does not rule out the
necessity for more precise calculations, but it is justified in view
of the present uncertainty related to such calculations. Its main
advantage is that of exhibiting explicitly the role of the lattice
symmetry, core states and lattice parameter on the energy band
structure of different compounds\dots''. Discussing the validity of
such a simple approach, they observed that ``\dots the critical role
played by the few lowest Fourier coefficients of the potential with
the corresponding terms from the core states explains the good
results obtained in spite of the poor knowledge of the crystal
potential''.

The main problem presented by the perturbative approach introduced
by Bassani and Celli was  that the Fourier coefficients of the
crystal potential were greater than the separations of the
electronic levels in the empty lattice:  for this reason a
perturbative method could not be applied. This point has been
discussed by Bassani in a private communication \cite{rossi}:
``...we noticed that the orthogonality condition of the inner states
introduced a correction in the potential that canceled its
contribution near the nucleus:  this fact justified the use of the
perturbation theory. J.C. Phillips and L. Kleinmann suggested that
the two terms should be considered as a pseudo~-~potential and that
its Fourier coefficients could be considered as available
parameters. [In the work published in 1961] we were able to prove
the existence of some degrees of freedom in defining the
pseudo~-~potential, due to the fact that the system formed by plane
waves and inner states was overcompleted. Thus, a general form of
the pseudo~-~potential could be developed with many practical
benefits. Among the works that used the pseudo~-~potential approach,
I believe that those were the most important and pioneering
contributions. We showed it was possible to neglect the Fourier
coefficients of the pseudo~-~potential corresponding to high values
of K of the reciprocal lattice because at short distances the
repulsive terms due to the inner states and the crystal potential
mutually cancel  themselves almost exactly. In such a way we
calculated the energy bands of Ge and GaAs by choosing the values of
three parameters and using an average potential V(0) adjusted to the
value of the energy gap. The accuracy of these calculations turned
out to be outstanding and the method based on  pseudo~-~potential is
still now used for semiconductors\dots''.

The OPW method, in the perturbation approximation introduced by
Bassani and Celli, was used by Bassani himself and Knox for
computing lowest conduction states in fcc solid Argon at the
symmetry points \cite{bassknox}. The valence and conduction states
were treated by two different approaches. For the valence states
they used the Tight-Binding approximation and for the conduction
band they applied the second-order perturbation approximation
developed by Bassani and Celli. This calculation was performed in
the one-electron approximation.

The many-body extension of OPW method was analyzed by Bassani and
co-workers in 1962 (the correlation term was introduced as a screen
effect of the exchange term) and these improvements allowed the
calculation of the band structures of  rare gases and alkali halides
\cite{bassrob}. In the following years Bassani published other
contributions on  band structure calculation. Among them, a paper
(by Bassani, Phillips and Brust) on the band structure of Germanium
\cite{bassbrust1}, where the reflectivity data of $Ge$ and $GaAs$
are analyzed by using the pseudo~-~potential technique. A similar
approach was used in the study of the effect of alloying and
pressure on the band structure of Silicon and Germanium
\cite{bassbrust2}.

Bassani and Yoshmine applied the OPW method to the zinc-blende
lattice \cite{bassyosh}. The calculation of valence and conduction
eigenvalues and eigenfunctions was carried out for a few  IV-group
elements and III-V compounds. The results obtained  by adopting
suitable approximations for the core states and exchange potential,
showed that all the investigated materials were semiconductors or
insulators.

In 1963 Bassani and Liu calculated the electronic energy band
structure including spin-orbit coupling effects for semiconducting
gray Tin \cite{bassliu}. The effect of pressure on the energy band
structure was also investigated. In this case, too, the calculated
energy gap turned out to be in agreement with the experimental data.
The correlation effects were taken into account also in the
calculation of the energy band structure in  AgCl and AgBr carried
out by Bassani, Knox and Fowler in 1965 \cite{bassknofow}. This
calculation demonstrated  that in these cases the maximum of the
valence band was positioned at point L in the Brillouin zone and the
minimum of the conduction band was at point K$=$0. The consequent
production of an indirect absorption in the visible spectrum and a
long life time for electrons and holes provided a first explanation
of the photographic properties of these crystals \cite{bassfowl-ph}.

The research on the energy band structure in AgCl and AgBr was
presented also at the SIF Meeting in 1964, together with another
work written by Bassani and Pastori on the symmetry properties of
electronic levels in layer compounds like GaS. The final paper on
the band structure calculation in the layer compounds was published
by Bassani and Pastori Parravicini in 1967  \cite{basspast}. The
band structure, calculated using a Tight-Binding semi~-~empirical
approach, was related to the basic properties of these compounds,
and a few features of the optical excitation spectrum were
discussed. The method used in this work represented a pioneering
application of the so-called ``semi-empirical tight binding
technique'' that is still now used to study complex systems and
nano~-~structures. The results concerning the electronic structure
of graphite explained also the electron energy loss spectrum, as
described in the work published in 1970 by Bassani and Tosatti
\cite{tosbass}.

Bassani dealt also with the electric field effects on optical
transitions \cite{bassaym} \cite{bassaym1} and the magnetic field
effects on energy band structure. The latter topic was discussed in
a paper published in 1967 with A. Baldereschi \cite{baldbass}, where
the existence of Landau levels in all critical points was
demonstrated.

Among the works carried out by Bassani in the 1960s one should not
forget the results concerning the correlation between band structure
and impurity states. In a paper  by Bassani, Iadonisi and Preziosi,
 the existence of a few
resonant states in the optical absorbtion region due to secondary
minima of the conduction band was investigated \cite{bassiadpre}.
Such a kind of resonances, described for the first time by the
Italian physicists, was experimentally confirmed in the following
years.
\\
\\
The researches carried out by Bassani strongly contributed to form
the first generation of Italian theorists in solid state physics.
The great number of studies carried out by Bassani, such as the ones
on the pseudo~-~potential developed with Celli and the
collaborations with several foreign scientists, brought his name to
the attention of the international community. Besides, the studies
published by Bassani stimulated the interest in semiconductor
physics in Italy during the Sixties.

\subsection{Other theoretical researches on semiconductors}
\label{sec:theo2} In the decade 1955$-$65 Italian theorists in the
field of condensed matter focused their attention on the following
topics: i) calculation of electronic level in solids; ii) study of
crystal potentials; iii) analysis of theoretical aspects related to
semiconductor devices.  Here, we shall deal only with researches on
semiconductors. For the other topics, the interested reader could
see \cite{rossi}.

In 1964, Manca and Aramu developed a model that allowed them to
explain
 the semiconducting phase of Fe$_{2}$Te$_{3}$ compound.
That work was first presented at the SIF Meeting and then published
in ``Il Nuovo Cimento''\cite{ara-tel}.

As previously underlined, the invention of the transistor
represented the spark that ignited a huge research on the technical
aspects of semiconductor physics. In Italy during the early 1960s,
such researches were mainly developed by Forlani and Minnaja working
at Olivetti (see the previous section). The activity of these two
scientists has been already discussed; we recall here the already
cited paper published in 1964 on the conduction mechanism in
Si$-$SiO$_{2}$$-$Al structures \cite{fm-cp}. The existence of charge
states predicted by this theoretical framework was confirmed by J.
Lindmajer a few years later \cite{lind}.

\section{Conclusions}
\label{sec:conc}

As shown in the introduction, Italian researchers who, in the
Fifties, began to work on condensed matter physics, or more
specifically, on semiconductors, had to overcome a knowledge gap of
about twenty five years. In the mid of the Sixties they had done
their job. The number of researchers in the field of solid state
physics was still low, if compared with that of the leading
countries. However, the cultural heritage left to the new
generations   by the pioneers we have talked about, allowed this new
generations of scientists to build their work on solid basis. The
pioneers had accomplished their {\em historical} task. In front of
this accomplishment, the evaluation of their work on the basis of
international standards belongs to a background scene. Anyway, for
completeness, we have reported in Table \ref{tab:4} the number of
citations of Italian scientists found in the series
\emph{Semiconductors and Semimetals} edited by  R.K. Willardson and
A.C. Beer and the series \emph{Solid State Physics} edited by F.
Seitz, D. Turnbull and H. Ehrenreich.
\begin{table}
\caption{Italian authors quotations in \emph{Semiconductors and
Semimetals} (SaS) and  \emph{Solid State Physics} (SSP)}
\label{tab:4}                    
\begin{tabular}{lcc}
\hline\noalign{\smallskip}
\textbf{Author}&\textbf{\emph{SaS}}&\textbf{\emph{SSP}}\\
\noalign{\smallskip}\hline\noalign{\smallskip} \hline \hline
Airoldi&-&2\\
Asdente&-&1\\
Bassani&19&3\\
Bertolotti&-&1\\
Bonfiglioli&5&2\\
Brovetto&5&2\\
Busca&1&1\\
Celli&3&-\\
Chiarotti&2&1\\
Della Pergola&-&1\\
Del Signore&1&-\\
Fieschi&-&3\\
Forlani&-&1\\
Frova&16&2\\
Fuhrman&-&1\\
Fumi&-&5\\
Germagnoli&-&2\\
Grassano&1&-\\
Grasso&-&1\\
Levialdi&1&1\\
Manca&1&-\\
Minnaja&-&1\\
Nucciotti&-&1\\
Palmieri&1&1\\
Papa&-&1\\
Rosei&1&-\\
Samoggia&1&1\\
Sette&-&2\\
Tosi&-&1\\
Vitali&-&1\\
Wanke&1&1\\
\hline
\textbf{Total}&59&40\\
\hline
\noalign{\smallskip}\hline
\end{tabular}
\end{table}
The total numbers of works quoted are about 5000 in
\emph{Semiconductors and Semimetals} and about 2800 in \emph{Solid
State Physics}. We leave to the reader the task of commenting the
table and of finding agreements or discrepancies with the feelings
suggested by the present paper. However, we can not refrain from the
warning that the citation game is often ruled by criteria that are
independent from the scientific value of the quoted or unquoted
papers.

\newpage
\section*{Acknowledgements}      
\label{sec:ack} Thanks to  Paolantonio Marazzini for a critical
reading of the manuscript and
 valuable suggestions; and to Giuseppe Giuliani for his continuous encouragement.

\end{document}